\documentclass[letterpaper]{article}

\usepackage{natbib,alifeconf}  %% The order is important

\title{Self-Organization and Artificial Life: A Review}
\author{Carlos Gershenson$^{1}$, Vito Trianni$^{2}$, Justin Werfel$^{3}$ \and Hiroki Sayama$^{4}$ \\
\mbox{}\\
$^1$Universidad Nacional Aut\'{o}noma de M\'{e}xico, Mexico City, Mexico\\
$^2$Institute of Cognitive Sciences and Technologies, Italian National Research Council, Rome, Italy\\
$^3$Wyss Institute for Biologically Inspired Enginering, Harvard University, Cambridge, MA 02138, USA\\
$^4$Binghamton University, Binghamton, NY 13902, USA\\
cgg@unam.mx, vito.trianni@istc.cnr.it,
justin.werfel@wyss.harvard.edu, sayama@binghamton.edu} % email of corresponding author

\begin{document}
\maketitle

\begin{abstract}
% Abstract length should not exceed 250 words
Self-organization has been an important concept within a number of disciplines, which Artificial Life (ALife) also has heavily utilized since its inception. The term and its implications, however, are often confusing or misinterpreted. In this work, we provide a mini-review of self-organization and its relationship with ALife, aiming at initiating discussions on this important topic with the interested audience. We first articulate some fundamental aspects of self-organization, outline its usage, and review its applications to ALife within its soft, hard, and wet domains. We also provide perspectives for further research.
\end{abstract}

\section{What is self-organization?}

The term ``self-organizing system'' was coined by \cite{Ashby1947sos} to describe phenomena where local interactions between independent elements lead to global behaviors or patterns. The phrase is used when an external observer perceives a pattern in a system with many components, and this pattern is not imposed by a central authority external to those components, but rather arises from the collective behavior of the elements themselves. Natural examples are found in areas such as collective motion \citep{Vicsek2012}, as when birds or fish move in flocks or schools exhibiting complex group behavior; morphogenesis \citep{Doursat2011}, in which cells in a living body divide and specialize to develop into a complex body plan; and pattern formation \citep{Cross1993} in a variety of physical and chemical systems, such as convection and crystal growth as well as the formation of patterns like stripes and spots on animal coats.

A formal definition of the term runs into difficulties in agreeing on what is a \emph{system}, what is \emph{organization}, and what is \emph{self} \citep{GershensonHeylighen2003a}, none of which are perfectly straightforward. However, a pragmatic approach focuses on when it is useful to describe a system as self-organizing \citep{GershensonDCSOS}. This utility typically comes when an observer identifies a pattern at a higher scale but is also interested in phenomena at a lower scale; there then arise questions of how the lower scale produces the observables at the higher scale, as well as how the higher scale constrains and promotes observables at the lower scale. For example, bird behavior leads to flock formation, and descriptors at the level of the flock can also be used to understand regulation of individual bird behavior \citep{KeysDugatkin1990}.

Self-organization has been an important concept within a number of disciplines, such as statistical mechanics \citep{Wolfram1983,Crutchfield2011}, supramolecular chemistry \citep{Lehn2017}, and computer science \citep{mamei2006case}. Artificial Life (ALife) frequently draws heavily on self-organizing systems in different contexts \citep{Aguilar2014The-Past-Presen}, starting in the early days of the field with studies of systems like snowflake formation \citep{Packard1986} and agent flocking \citep{reynolds87flocks}, and continuing to the present day. However, there are often confusion and misinterpretation involved with this concept, possibly due to an apparent lack of recent systematic literature. In this work, we aim at providing a mini-review of self-organization within the context of ALife, with a goal to open discussions on this important topic to the interested audience within the community. We first articulate some fundamental aspects of self-organization, outline ways the term has been used by researchers in the field, and then summarize work based on self-organization within \emph{soft} (simulated), \emph{hard} (robotic), and \emph{wet} (chemical and biochemical) domains of ALife. We  also provide perspectives of further research.

\section{Usage}

Ashby coined the term ``self-organizing system'' to show that a machine could be strictly deterministic and yet exhibit a self-induced change of organization \citep{Ashby1947sos}. This notion was further developed within cybernetics \citep{vonFoerster1960,Ashby1962}. In many contexts, a thermodynamical perspective has been taken, where ``organization'' is viewed as the opposite of entropy \citep{NicolisPrigogine1977}. Since there is an equivalence between Boltzmann-Gibbs entropy and Shannon information, this notion has also been applied in contexts related to information theory \citep{Fernandez2013Information-Mea}. In this view, a self-organizing system is one whose dynamics lead it to decrease its entropy or increase its information content. In the meantime, there are several other definitions of self-organization as well. For example, \cite{Shalizi2001} defines self-organization as an increase in statistical complexity, which in turn is defined as the amount of information required to minimally specify the state of the system's causal architecture. As an alternative to entropy, the use of the mean value of random variables has also been proposed \citep{Holzer:2011}.

The recent subfield of \emph{guided self-organization} explores mechanisms by which self-organization can be regulated for specific purposes---that is, how to find or design dynamics for a system such that it will have particular attractors or outcomes \citep{Prokopenko:2009,Ay2012Guided-self-org,GSO2013,GSOInception2014,ProkopenkoGershenson2014}. Much of this research is based on information theory. For example, the self-organization of random Boolean networks \citep{Kauffman1969,Kauffman1993} can be guided to specific dynamical regimes \citep{Gershenson:2010}. The concept of self-organization is also heavily used in organization science, with relevance to artificial society models \citep{gilbert1995artificial,EpsteinAxtell1996}.

While there may be no single agreed-on definition of self-organization, this lack need not be an insurmountable obstacle for its study, any more than a lack of a unanimous formal definition of ``life'' has been an obstacle for progress in the fields of biology or ALife. In what follows, we provide a concice review of how self-organization has contributed to the progress of ALife.

\section{Domains}

One way to classify ALife research is to divide it into \emph{soft}, \emph{hard}, and \emph{wet} domains, roughly referring to computer simulations, physical robots, and chemical/biological research (including living technology as the application of ALife \citep{Bedau:2009}), respectively. Self-organization has played a central role in work in all three domains.

\subsection{Soft ALife}

Soft ALife, or mathematical and computational modeling and simulation of life-like behaviors, has been linked to self-organization in many sub-domains.  Cellular automata (CAs) \citep{Ilachinski2001}, one of the most popular modeling frameworks used in earlier forms of soft ALife, are well-explored, illustrative examples of self-organizing systems. A CA consists of many units (cells), each of which can be in any of a number of discrete states, and each of which repeatedly determines its next state in a fully distributed manner, based on its current state and those of its neighbors. With no central controller involved, CAs can spontaneously organize their state configurations to demonstrate various forms of self-organization: dynamical critical states such as in sand-pile models \citep{BakTangWiesenfeld1988} and in the Game of Life \citep{BakChenCreutz1989}, spontaneous formation of spatial patterns \citep{Young1984,Wolfram1984,ErmentroutEdelsteinKeshet1993}, self-replication \footnote{Note that earlier literature on self-reproducing cellular automata \citep{vonNeumann1966,Codd1968} is not included here, because those models typically had a clear separation between a central universal controller and a structure that is procedurally constructed by the controller; thus they may not constitute a good example of self-organization as discussed in this article.} \citep{Langton1984,Langton1986,ReggiaEtAl1993,Sipper98}, and evolution by variation and natural selection \citep{Sayama1999,Sayama2004,SalzbergSayama2004,SuzukiIkegami2006,OrosNehaniv2007,OrosNehaniv2009}. Similarly, partial differential equations (PDEs), a continuous counterpart of CAs, have an even longer history of demonstrating self-organizing dynamics \citep{Turing1952,GlansdorffPrigogine1971,FieldNoyes1974,Pearson1993}.

Another representative class of soft ALife that shows self-organization comprises models of collective behavior of self-driven agents \citep{Vicsek2012}. Reynolds' Boids model \citep{reynolds87flocks} is probably the best known in this category. In this work, self-propelled agents (``boids'') move in a continuous space according to three kinetic rules: cohesion (to maintain positional proximity), alignment (to maintain directional similarity), and separation (to avoid overcrowding and collision). A variety of related models have since been proposed and studied, including simplified, statistical-physics-oriented ones \citep{VicsekEtAl1995,LevineEtAl2000,AldanaEtAl2007,NewmanSayama2008} and more detailed, behavioral-ecology-oriented ones \citep{Couzin:2002,KunzHemelrijk2003,HildenbrandtEtAl2010}. These models produce natural-looking flocking/schooling/swarming collective behaviors out of simple decentralized behavioral rules, and they also exhibit phase transitions between distinct macroscopic states.

Such collective behavior models have been brought to \emph{artificial chemistry} studies \citep{Dittrich2001Artificial-Chem,BanzhafYamamoto2015} as well, such as \emph{swarm chemistry} and its variants \citep{Sayama2008Swarm-Chemistry,KreyssigDittrich2011,Sayama2011,Sayama2012,ErskineHerrmann2015}, in which kinetically and chemically distinct species of idealized agents interact to form nontrivial spatiotemporal dynamic patterns. More recently, these collective behavior models have also been actively utilized in \emph{morphogenetic engineering} \citep{Doursat2011,DoursatEtAl2012}, in which researchers attempt to achieve a successful merger of self-organization and programmable architectural design, by discovering or designing agent rules that result in specific desired high-level patterns.

Other examples of self-organization in soft ALife are found in simulation models of artificial societies. Their roots can be traced back to the famous segregation models developed by Sakoda and Schelling back in the early 1970s \citep{Sakoda1971,Schelling1971,Hegselmann2017}, in which simple, independent decision making by individual agents would eventually cause a spatially segregated state of society at a macroscopic level. Agent-based simulation of artificial societies has been one of the core topics discussed in the ALife community \citep{epstein1996,Lansing2002}, and has elucidated self-organization of issues in social order such as geographical resource management \citep{LansingKremer1993,Bousquet2004313}, cooperative strategies \citep{LindgrenNordahl1993,Brede2011,AdamiEtAl2016,IchinoseSayama2017}, and common languages \citep{Steels1995,Kirby2002,SmithEtAl2003,LipowskaLipowski2012}. The literature on self-organization of adaptive social network structure \citep{gross2009adaptiveNets,BrydenEtAl2010,GeardBullock2010} may also be included in this category.

\subsection{Hard ALife}

Robots can be considered to be life-like artifacts in their ability to sense their physical environment and take action in response. Physical agents, even very simple ones, can evoke in the observer a particularly strong sense of being animate. From W. Grey Walter's tortoises \citep{walter1950,walter1951machine}, to simple machines based on the principles of Braitenberg's vehicles \citep{Braitenberg:1986}, to other reactive robots \citep{Brooks1989}, to recent biomimetic and bioinspired designs \citep{SaranliEtAl2001,WoodEtAl2013,KimWensing2017}, building artificial life as physically embodied hardware allows it to exploit the rich dynamics underlying the interaction between itself and its environment, so that even simple mechanisms and behavioral rules can confer sophisticated life-like attributes to limited machines \citep{Simon1969}. Still higher complexity can be attained either by increasing the sophistication of a single robot, or by increasing the number of robots in a system that, through the resulting interaction and self-organization, can then evince more sophisticated abilities collectively, from adaptive responses to group decision making. 

Physical hardware has the strong advantage that the physical characteristics of the system (dynamics, sensor performance, actuator noise profiles, etc.) are by definition realistic, whereas simulations are necessarily simplified and typically fail to capture phenomena that only become evident through material experimentation \citep{BrooksMataric1993,RubensteinEtAl2014}. Conversely, while simulation can readily handle very large numbers of agents, hardware considerations (cost, space, scalability of operation, etc.) have traditionally limited hard ALife studies to using a small number of robots. In some scenarios, self-organizing phenomena of interest do not necessarily require a large number of robots; when the mechanism for coordination is based on \emph{stigmergy} (persistent information left in a shared environment), the important element is a large number of interactions between robot and environment, and even a single robot could suffice \citep{BeckersEtAl2000,WerfelEtAl2014}. More recently, hardware advances have made it possible to conduct physical experiments with robots in numbers exceeding a thousand \citep{RubensteinEtAl2014}. 

Physical experiments have been used to explore self-organizing phenomena in a variety of areas. Aggregation of objects has been studied from a physics perspective \citep{GiomiEtAl2013}; in ways inspired by behavior observed in living systems, such as cockroaches or bees \citep{GarnierEtAl2008,KernbachEtAl2009}; and using controllers designed through automatic methods like artificial evolution \citep{DorigoEtAl2004,FrancescaEtAl2014}. Another topic is collective navigation, in which groups of robots coordinate their overall direction of motion and collectively avoid obstacles \citep{BaldassarreEtAl2007,TrianniDorigo2006,TurgutEtAl2008}.  In other collective decision-making processes, positive feedback from recruitment processes and negative feedback from cross-inhibition contribute to shape the outcome \citep{ReinaEtAl2018,ValentiniEtAl2015,ScheidlerEtAl2016,GarnierEtAl2009,GarnierEtAl2013,KernbachEtAl2009,FrancescaEtAl2014,ValentiniEtAl2017}.
Self-assembly \citep{Whitesides2002} is another form of self-organization, with several examples in hard ALife of self-assembling or self-reconfiguring robots \citep{MurataEtAl1994,HollandMelhuish1999,StoyNagpal2004,ZykovEtAl2005,DorigoEtAl2006,AmpatzisEtAl2009,RubensteinEtAl2014}.

\subsection{Wet ALife}

Wet ALife, or physico-chemical synthesis of life-like behaviors, extensively utilizes self-organization as its core principle. A classic example is the spatial pattern formation in experimentally realized reaction-diffusion systems, such as the Belousov-Zhabotinsky reaction \citep{vanag2001pattern,adamatzky2008universal} and Gray-Scott-like self-replicating spots \citep{lee1994experimental}, where dynamic patterns self-organize entirely from spatially localized chemical reactions. Similar approaches can also be taken by using microscopic biological organisms (e.g., slime molds) as the media of self-organization \citep{adamatzky2008universal,adamatzky2015would}. In the origins of life research, molecular self-assembly plays the essential role in producing protocell structures and their metabolic dynamics \citep{rasmussen2003bridging,hanczyc2003experimental,rasmussen2004transitions,Protocells2008}.

More recently, dynamic behaviors of macroscopically visible chemical droplets, a.k.a. \emph{liquid robots} \citep{vcejkova2017droplets}, has become a focus of active research in ALife. In this line of research, interactions among chemical reactions, physical micro-fluid dynamics and possibly other not-yet-fully-understood microscopic mechanisms cause self-organization of spontaneous movements \citep{hanczyc2007fatty,cejkova2014dynamics} and complex morphology \citep{vcejkova2018multi} of those droplets. Moreover, droplet-based systems have also been used to demonstrate artificial evolution in experimental chemical systems \citep{parrilla2017adaptive}.

\section{Perspectives}
\label{sec:disc-concl}

As already mentioned above, we can understand a self-organizing system as one in which organization increases in time. However, it can be shown that, depending on how the variables of a system are chosen, the same system can be said to be either organizing or disorganizing \citep{GershensonHeylighen2003a}. Moreover, in several examples of self-organization, it is not straightforward to identify the \emph{self} of the system, as oftentimes all elements composing the system can be ascribed equal agency. Finally, in cybernetics and systems theory, the dependency of the boundaries of a system on the observer has thoroughly been discussed \citep{Gershenson2013The-Past-Presen}: one wants to have an objective description of phenomena, but descriptions are necessarily made by observers, making them partially subjective.
It becomes clear, then, that discussing self-organization requires the identification of what is \emph{self} and what is \emph{other}, and what are the elements that are increasing in their \emph{organization}. Similar issues have been tackled by \cite{MaturanaVarela1980} in the definition of living systems as autopoietic systems. According to this tradition, a living system is inherently self-organising because the \emph{self} is continuously produced or renewed by processes brought forth by the system's internal components. In other words, an autopoietic system can be recognized as a unity with boundaries that encompass a number of simpler/elementary components that are at the basis of the organization of the system, as they are responsible for the definition of the system boundaries and for the (re)production of the very same components \citep{varela1974autopoiesis}. This is a peculiar characteristic of living systems. If life is deeply rooted in self-organization, so should be ALife, and the several acceptations of ALife discussed above demonstrate the richness of the links it holds with self-organization.

Looking at the perspectives of ALife, it can be useful to think of self-organization as the common language that unifies the soft, hard and wet domains. The ALife community can progress owing to shared concepts and definitions, and despite the mentioned difficulties, self-organization stands as a common ground on which to build shared consensus. Most importantly, we believe that the identification and classifications of the \emph{mechanisms} that underpin self-organization can be extremely useful to synthesize novel forms of ALife and gain a better understanding of life itself.
These mechanisms should be identified at the level of the system components and characterized for the effects they have on the system organization. Mechanisms pertain to the modalities of interaction among system components (e.g., collisions, perceptions, direct communication, stigmergy), to behavioural patterns pertaining to individual components (e.g., exploration vs.\ exploitation), and to information enhancement or suppression (e.g., recruitment or inhibitory processes). The effects of the mechanisms should be visible in the creation of feedback loops---positive or negative---at the system level, which determine the complex dynamics underlying self-organization.
We believe that, by identifying and characterising the mechanisms that support self-organization, the synthesis of artifacts with life-like properties would be much simplified. In this perspective, mechanisms underlying self-organization can be thought of as \emph{design patterns} to generate artificial lives \citep{Babaoglu:2006hy,FernandezMarquez:2012es,Reina:2014uc}. By exploiting and composing them, different forms of ALife could be designed with a principled approach, owing to the understanding of the relationship between mechanisms and system organization.

The possibility of exploiting self-organization for design purposes is especially relevant toward the development of \emph{living technologies}, that is, technologies presenting features of living systems \citep{Bedau:2009}, such as robustness, adaptability, and self-organization, which can include self-reconfiguration, self-healing, self-management, self-assembly, etc., often named together as ``self-*'' in the context of autonomic computing \citep{Poslad2009}.
Self-organization has been used directly in living technologies within a variety of domains \citep{Bedau2013IntroductionLT}, from protocells \citep{Protocells2008} to cities \citep{Gershenson:2013}, and also several methodologies that use self-organization have been proposed in engineering \citep{Frei:2011}. A major leap forward can be expected when principled design methodologies are laid down, and a better understanding of self-organization for ALife can be at the forefront of the development of such methods.

% When self-organization is NOT useful? 
% Embodied cognition
% Evolutionary algorithms
% (but open-ended evolution)
As a final discussion, and as food for thought, it is also worth considering when self-organization is \emph{not} useful in the context of ALife. Tracing a clear line across the domain is of course impossible, but our reasoning above provides some suggestions. Indeed, self-organization does not account for every life-like process, for instance when there is no clear increase in organization. For instance, hard ALife has strongly developed the concept of embodied cognition and morphological computation \citep{Pfeifer:2009ba}, where the dynamics of mind-body-environment interaction are fundamental aspects. These dynamics, albeit very complex, are not easily described within the framework of self-organization.
Evolution is also very much represented within ALife, but typical generational evolutionary algorithms do not present clear elements of self-organization, as the progression in organization through generations is mediated by a central authority that defines selection of the fittest individuals. Open-ended evolution \citep{taylor2016open} makes a difference when such a centralized authority does not exist and progress is observable in mixing populations of interacting individuals. Exploring possibilities of infusing self-organization into those processes that were traditionally not oriented toward self-organization would be potentially a very fruitful direction of research.
%
%\texttt{Other issues?}

%\texttt{Closing remarks...}

Needless to say, this mini-review we presented here is not meant to be a complete, comprehensive review of self-organization and ALife, given the limitation of space. We plan to expand the work substantially in the near future to develop a more thorough review for publication elsewhere.

% Lots of balances... order/chaos, robustness/adaptability, production/destruction, exploration/exploitation... self-organization can be useful to let systems find the appropriate balance for their current context, as it is known that the optimal balance can change (Gershenson and Helbing, 2015)

% Future Perspectives

% How can one make self-organization programmable?
% Can one predict the macroscopic outcomes of self-organization?
% What are the roles of self-organization in ALife challenges (e.g., open-ended evolution)?

%\section{Acknowledgements}

%This work was supported by NSF grant No.\ PHY-9723972.

\footnotesize
\bibliographystyle{apalike}
\bibliography{refs,carlos,justin,hiroki-additional} % replace by the name of your .bib file

\end{document}